# Overton Engage: A Structured Database and Matching System for Academic Policy Engagement Opportunities

Ceire Wincott, Angel Luis Jaso Tamame, Susan Collard, Euan Adie, Katie Shamash

**Abstract**

Academic policy engagement, the structured processes through which researchers contribute evidence and expertise to public decision-making, is shaped not only by research quality but by the accessibility of engagement opportunities. In practice, these opportunities are fragmented across institutions and platforms, unevenly advertised, and difficult to discover systematically (Parker et al., 2022), limiting both individual participation and comparison. We present Overton Engage (https://app.overton.io/ui/opportunities), a structured database of publicly documented academic policy engagement opportunities, together with a semantic matching system that links opportunities to researchers based on similarity between opportunity descriptions and publication records. We characterise the composition of the database across policy domains, countries, and opportunity types, and present UK-focused analyses comparing engagement opportunity topics with published policy documents. We further demonstrate an illustrative comparison of consultation topics between the UK and Australia, and apply a matching system to assess how closely research produced by UK higher education institutions aligns, topically, with domestic policy opportunities.

Our results suggest that publicly documented engagement opportunities are unevenly distributed across policy domains and countries, though this may reflect collection bias. Matching analyses reveal a positive relationship between institutional publication volume and high-confidence match rates, but also that research specialisation can compensate for lower output volume in specific policy domains.

The database itself is freely available and we welcome collaboration from researchers, policymakers, and institutions.

**Introduction**

Academic policy engagement refers to structured interactions through which researchers contribute evidence and expertise to public decision-making processes, including formal and informal consultations, advisory roles, commissioned research, and related forms of policy-facing activity (Oliver & Cairney, 2019). Rather than operating as a simple transfer of findings into policy, research use occurs through multiple pathways shaped by institutional context and access to engagement mechanisms (Weiss, 1979; Nutley et al., 2007).

Early institutional models in both the USA and the UK positioned publicly supported research as relevant to government capacity and societal needs with an explicit role in informing government action and addressing societal challenges (Bush, 1945; Haldane, 1918). Subsequent scholarship in both the United States and the United Kingdom shows that the effectiveness of academic-policy engagement depends not only on research quality, but also on the institutional structures and channels through which engagement opportunities are surfaced and accessed, such as targeted calls for evidence, formal fellowship schemes, or nominations to advisory bodies (National Research Council, 2012; Cairney, 2016). Although formal mechanisms such as fellowships, advisory committees, and commissioned research exist, participation in these processes is frequently shaped by informal networks, prior relationships, and uneven access to information about these opportunities (National Research Council, 2012; Oliver & Cairney, 2019). As a result, engagement opportunities may systematically favour well-connected researchers and institutions, limiting the diversity of perspectives that contribute to policy processes.

Comparative and international studies indicate substantial variation in how governments engage academic expertise. These differences include the extent to which engagement opportunities are publicly advertised versus invitation-based, the degree of transparency surrounding engagement processes, and the consistency with which opportunities are documented and made accessible (Court & Young, 2006; Boaz et al., 2019; OECD, 2020). Consequently, policy engagement opportunities are more visible and systematically discoverable in some jurisdictions than in others, limiting cross-national comparability and constraining the generalisability of analyses based on publicly available data.

We present Overton Engage ([https://engage.overton.io](https://engage.overton.io)), comprising two systems designed to reduce informational and practical barriers to academic policy engagement by improving the discoverability and accessibility of publicly available opportunities worldwide: (i) a structured database of publicly documented academic policy engagement opportunities, and (ii) a matching system that links opportunities to researchers based on semantic similarity between opportunity descriptions and publication records.

Overton Engage is directly inspired by the ARI Database ([https://ari.org.uk](https://ari.org.uk)), which pulls together information on Areas of Research Interest (ARIs) from central government departments in the UK, but differs by having a broader scope (collecting opportunities from all countries, not just the UK, and a range of different opportunity types).

To support personalised discovery, a semantic matching system is used to link opportunities to researchers based on the similarity between their publication records and opportunity descriptions.. This system is intended to illustrate how the database can be operationalised rather than to provide an exhaustive evaluation of matching performance (Fortunato et al., 2018; Hicks et al., 2015). Full details of the matching procedure are described in the Methods section.

This paper has three components. First, we characterise the Overton Engage database, describing how collected opportunities are distributed across policy domains, countries, and engagement types. Second, we compare UK engagement opportunities with published UK policy documents to assess the relationship between policy demand and policy output. Third, we use a matching system to analyse how the research focus of UK higher education institutions correlates with the topics of publicly advertised policy opportunities, both overall and within specific policy domains.

Together, these analyses illustrate what becomes possible when engagement opportunities are collected, standardised, and made systematically discoverable.

**Methods**

**The Overton Engage Database**

Overton Engage centralises and standardises information on policy engagement opportunities, reducing the time and effort required to identify relevant opportunities and enabling systematic search and comparison across policy organisations, topics, and countries. The data model captures key attributes, including requesting organisation, country, opportunity type, background context, contact details, and links to source material, supporting consistent interpretation across a varied range of sources. Opportunities are automatically classified to a COFOG (Classification of the Functions of Government) category, an internationally used classification system developed by the OECD and United Nations to group government activities and expenditures by functional policy domain ([https://unstats.un.org/unsd/classifications/cofog/revision](https://unstats.un.org/unsd/classifications/cofog/revision)). Information within the database is curated to support accessibility and use by an academic audience, with attention to clarity, consistency, and sufficient contextual detail to support informed engagement.

A fine-tuned embeddings model and semantic similarity between opportunities and academic publications are used to identify matches between researchers and engagement opportunities. Both publication abstracts and opportunity descriptions are converted into high-dimensional numerical representations (embeddings) that capture their semantic content, enabling similarity to be measured mathematically.

For each opportunity, the algorithm identifies the most semantically similar publication abstracts. The semantic similarity reflects the likelihood that a given publication and its authors may be able to contribute to that specific opportunity - the embeddings model having been trained to reflect this during a fine-tuning process using a large set of manually annotated training data.

Each opportunity / publication pair is classified as either "green" (a high confidence match), "orange", "yellow" or "red" (a low confidence match) based on the L2 distance between the embeddings of a publication abstract and an opportunity, with lower values indicating closer semantic alignment and thus higher confidence.

Researchers are then ranked according to both the frequency with which their works appear in this set and the relevancy scores of those works.

Whilst the database includes opportunities from multiple countries, coverage is constrained by the availability of publicly posted opportunities and by differences in publication and documentation practices. At present, coverage is substantially more complete for the United Kingdom than for most other jurisdictions. Accordingly, this paper treats the UK as the primary case for detailed characterisation and analysis, while presenting other countries descriptively and without claims of completeness.

**Institutional-level analysis**

For the matching analysis, 161 UK higher education institutions were processed sequentially. For each institution, the full publication corpus was matched against all opportunities in the Overton Engage database, and the number of matches in each confidence tier (Green, Yellow, Orange, Red) was recorded per opportunity. Results were then aggregated at the institution level. The percentage of high-confidence matches reported for each institution reflects the proportion of opportunities for which at least one publication from that institution achieved a Green match (L2 ≤ 0.288). Publication volume was measured as the total number of qualifying works published by each institution over the five-year window. Where analyses are restricted to specific COFOG categories (e.g. Health, Defence), only opportunities classified under the relevant category were included in the calculation.

**Results & Discussion**

**Data Characterisation**

We first describe the opportunity corpus assembled in Overton Engage as context for the system's downstream use. We then present UK-focused analyses that connect opportunity topics to policy outputs and to researcher-opportunity matching.

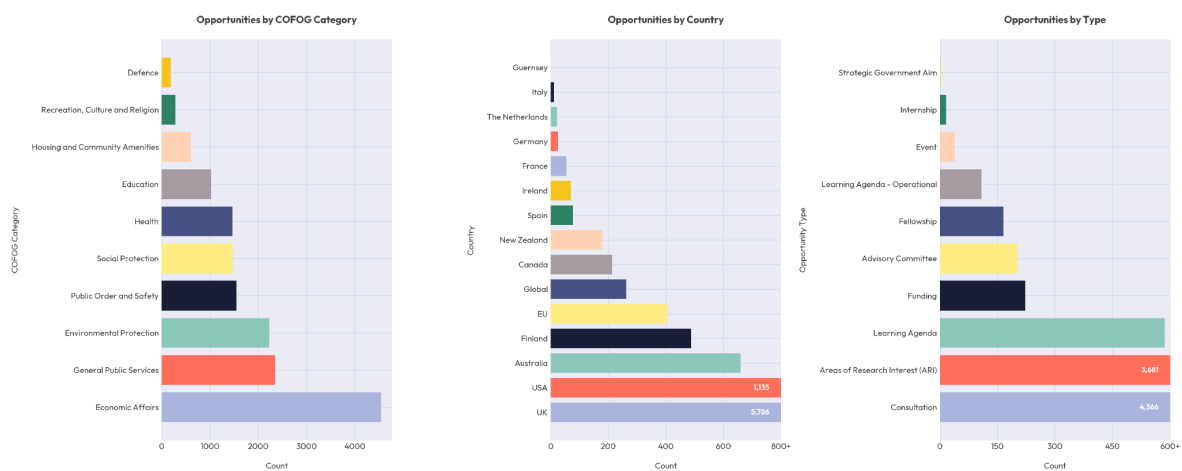

**Figure 1. Characterisation of the Overton Engage opportunity corpus:** (a) Distribution of opportunities across COFOG policy categories. (b) Distribution of opportunities by country. (c) Distribution of opportunities by opportunity type. Counts reflect the number of publicly available policy engagement opportunities collected and normalised in Overton Engage at the time of analysis. Distributions represent data availability rather than comprehensive national engagement activity.

Figure 1 summarises the composition of the Overton Engage opportunity corpus by COFOG category, country, and opportunity type.

Opportunities are unevenly distributed across COFOG categories (Figure 1a). Economic Affairs dominates, followed by General Public Services and Environmental Protection, while Education, Housing and Community Amenities, Recreation, Culture and Religion, and Defence are comparatively fewer.

Geographic coverage is strongly skewed towards a small number of countries (Figure 1b). The United Kingdom accounts for the majority of collected opportunities, with substantially smaller contributions from other countries. This skew likely reflects variation in language, publication practices, and current data collection coverage bias, rather than necessarily representing true differences in levels of academic-policy engagement activity.

In addition to policy-domain classification, each record is assigned an *opportunity type*, that captures the primary mechanism through which policymakers invite engagement with academic researchers. Opportunity types include ARIs, Consultations, Learning Agendas, Fellowships, Internships and Events. These categories are assigned during curation based on the stated purpose and structure of the opportunity, and are intended to support filtering, comparison, and aggregate analysis rather than to capture the full nuance of individual engagement processes.

Opportunity types are also unevenly represented (Figure 1c). ARIs and Consultations dominate the corpus, with Learning Agendas forming a smaller but notable category. Other engagement instruments account for a minority of records.

Together, these patterns describe how publicly documented policy engagement opportunities are distributed within the Overton Engage dataset, while highlighting that the dataset reflects only a subset of publicly visible policymaker-researcher interactions.

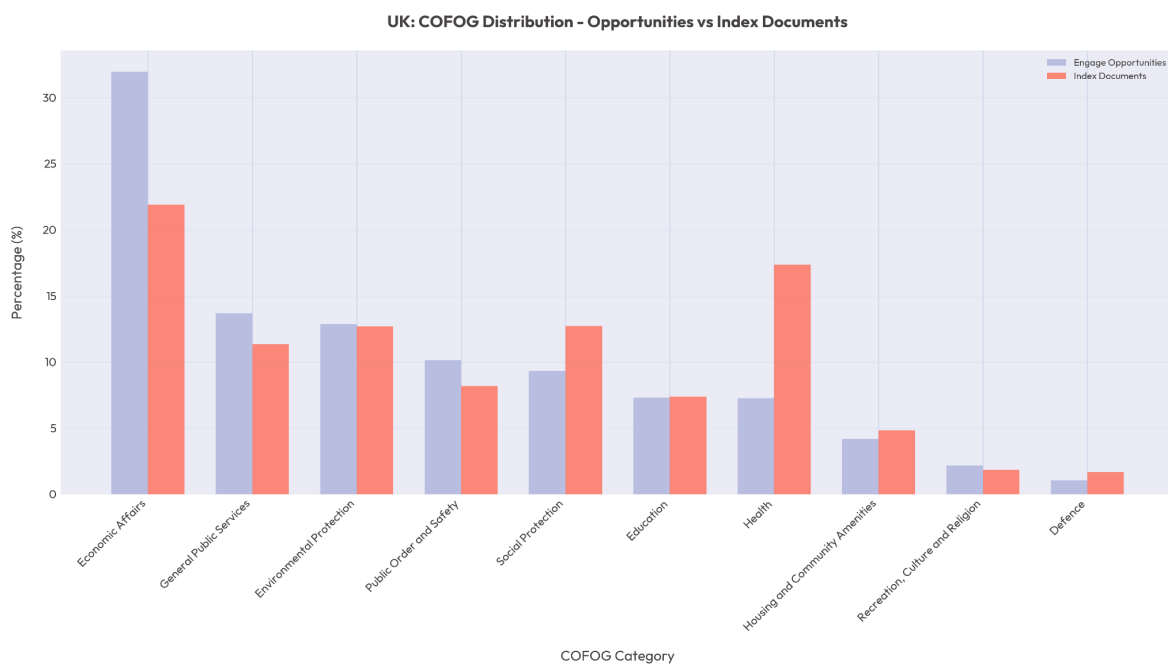

**Figure 2. UK COFOG distribution of engagement opportunities and published policy documents:** Percentage distribution of UK engagement opportunities in Overton Engage (blue) and UK policy documents which cite something in Overton Index (red) across COFOG policy categories. Percentages are calculated within each dataset and reflect relative topic emphasis rather than absolute volume.

Figure 2 compares the distribution of UK engagement opportunities in Overton Engage with the distribution of UK policy documents indexed in Overton Index, since 2023, across COFOG policy categories. Overton Index is a large-scale database of policy documents from global sources (Szomszor & Adie, 2022), including government departments, public bodies, and policy-oriented organisations such as think tanks, covering reports, strategies, guidance, and other materials intended to inform or reflect public decision-making. For this analysis, the policy document corpus is restricted to documents that cite at least one other research or policy document, in order to focus on materials that draw on external evidence. In contrast to these completed or ongoing policy outputs, Overton Engage captures explicit invitations for academic input, instances where policymakers actively seek external expertise rather than publish findings derived from it.

While both datasets draw from overlapping institutional ecosystems, the materials they contain differ in form and function. Policy documents represent completed or ongoing policy-related outputs, which vary substantially in purpose and format, ranging from high-level strategy documents to routine statistical or monitoring publications. Engagement opportunities, by contrast, are more tightly curated, reflecting explicit invitations for external contribution. These differences in document type and curation may account for some of the observed variation in topical distributions.

Overton Engage captures policy demand at the point of consultation, while the Index documents reflect published policy outputs from the same period. Contributions made through engagement opportunities may therefore not yet be visible in published policy documents. The comparison should accordingly be interpreted as contrasting contemporaneous policy demand with policy output over the same period, rather than as a direct measure of policy influence.

Overall, these results suggest broad alignment between the topics governments publish policy on and those for which they most frequently seek external input, with both datasets showing Economic Affairs and General Public Services as dominant categories. The most notable divergence is in Health, which is substantially more represented in published policy outputs than in engagement opportunities, and in Economic Affairs, which is more prominent in engagement opportunities than in published documents. This likely reflects a difference in volume of published engagement opportunities rather than any systematic difference in how health policy engages with research.

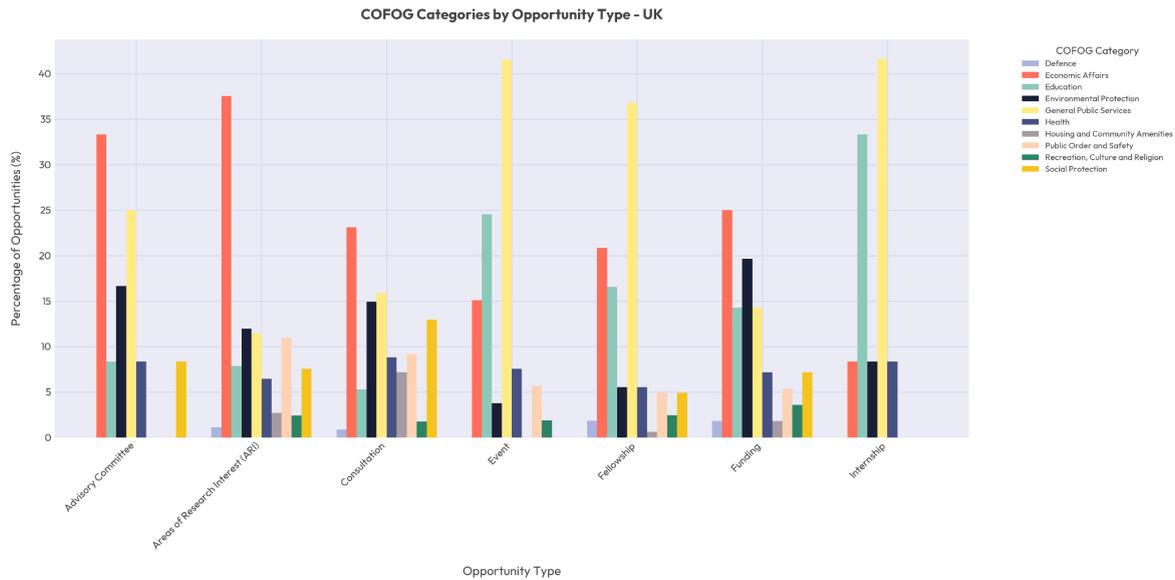

**Figure 3. Distribution of COFOG policy categories by opportunity type in the UK:** Percentage of UK engagement opportunities within each opportunity type assigned to each COFOG policy category. Percentages are calculated within opportunity types and illustrate how different engagement instruments are used across policy domains.

Different types of engagement opportunities may serve distinct functions in the policy process. Consultations are typically reactive instruments through which governments seek input on specific legislative or regulatory proposals, whereas ARIs represent a more proactive attempt to signal longer-term research priorities (Oliver et al., 2022). Fellowships and internships embed researchers directly within government, while funding opportunities and advisory committees reflect more sustained forms of academic-policy engagement.

If these mechanisms serve different functions, we might expect them to draw on different policy domains — for example, ARIs may be more concentrated in areas where governments are actively building an evidence base, while consultations may be more evenly distributed across domains reflecting the breadth of legislative activity.

Figure 3 shows how COFOG policy categories are distributed across engagement opportunity types in the UK. Clear differences emerge across opportunity types. ARIs are concentrated in Economic Affairs, followed by Health and Social Protection, reflecting their agenda-setting role. Consultations show a broader distribution across Economic Affairs, Public Order and Safety, General Public Services, and Social Protection. Together, these patterns indicate that different engagement instruments tend to be used more frequently in certain policy domains than in others, suggesting variation in how policymakers invite academic input across policy areas.

The reasons for this variation are not immediately clear from the data alone, but several explanations are plausible. One possibility is that certain policy domains are more culturally accustomed to structured academic engagement (Verboom & Baumann, 2022) — health policy, for instance, has a well-developed tradition of evidence-informed policymaking, which may explain why it features strongly in fellowships and funding opportunities that embed researchers in government. Economic affairs, by contrast, may be more amenable to the kind of broad, open-ended research signalling that ARIs represent.

Future work could examine whether these patterns reflect deliberate strategic choices by government departments or emerge organically from existing academic-policy networks.

**Comparing Consultations in the UK & Australia**

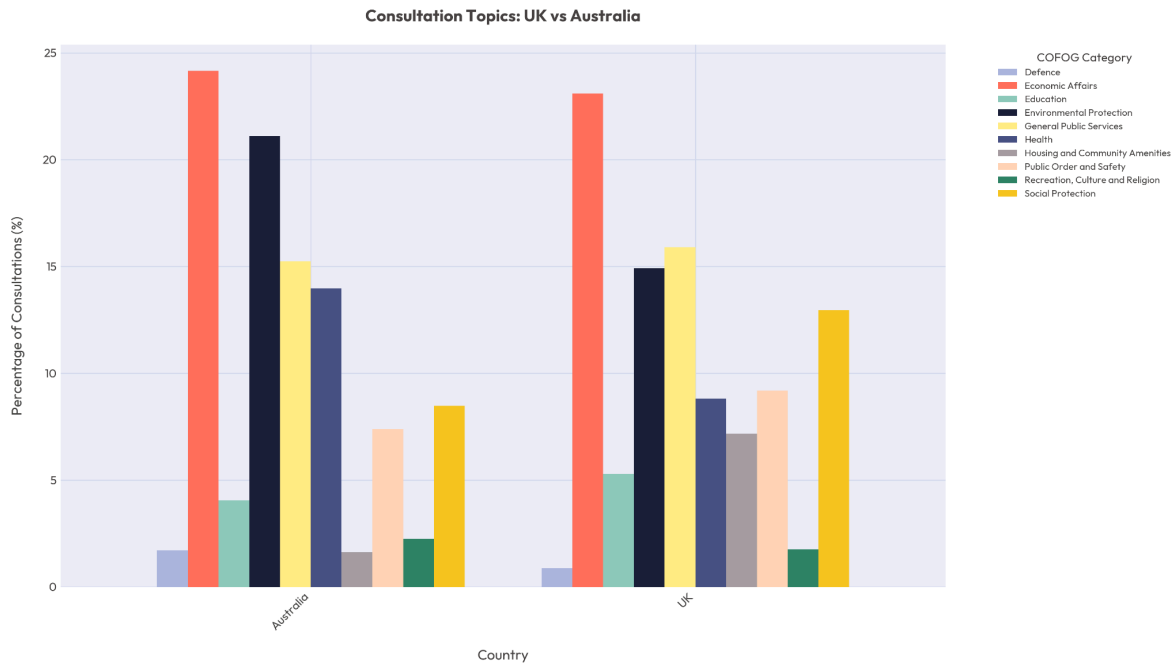

**Figure 4. COFOG distribution of consultation topics in the UK and Australia:** Percentage of consultation opportunities in each country assigned to COFOG policy categories. Percentages are calculated within each country's consultation corpus.

Comparison of consultation topics offers a way to assess whether the policy domains that attract engagement activity are specific to particular national contexts or reflect broader shared priorities. While a full multi-country comparison is beyond the scope of this paper, Figure 4 presents an illustrative comparison between the UK and Australia to demonstrate the type of analysis that becomes possible with this data and to situate the UK findings within a broader comparative frame.

Figure 4 compares the distribution of consultation topics across COFOG categories in the UK and Australia. In both countries, consultations are defined as formal processes through which governments or public bodies invite written evidence or feedback from experts and stakeholders on proposed policies, legislation, or strategic priorities. In both datasets, the majority of consultations originate from government departments, providing a broadly comparable institutional basis for analysis. The UK dataset includes a higher proportion of consultations from local government bodies than the Australian data.

This comparison is intended to highlight similarities and differences in how policy priorities are expressed through consultation activity across the two national policy systems, rather than to provide a definitive measure of overall consultation volume. Examining topic distributions therefore offers insight into how engagement is oriented across policy domains within different governance contexts.

A higher proportion of Australian consultations concern Health and Environmental Protection, while UK consultations have a higher proportion of General Public Services, Social Protection, and Housing and Community Amenities. These differences may reflect variation in national policy agendas, institutional responsibilities across levels of government, and the timing of major policy initiatives in each country, as well as differences in how consultation is used as a policy tool across domains. As such, the observed patterns should be interpreted as indicative of differing policy emphases rather than as direct measures of underlying societal need or problem severity.

Despite these differences, overall profiles are broadly similar, suggesting that consultations in both countries concentrate on a shared set of core policy domains. At the same time, national context shapes which topics are considered suitable for public consultation. As in earlier analyses, these results reflect only publicly advertised consultations captured in Overton Engage.

**Analysis of University-Opportunity Matching**

We next examine how these opportunities align with UK research capacity using the researcher matching system. To examine how well-positioned UK institutions are to engage with policy opportunities, institutional research output was matched against opportunities in the Overton Engage database using the system described in the Methods section. Publications associated with a compiled list of UK higher education institutions (see Supplementary Material) were retrieved via OpenAlex for the period 2020-2025.

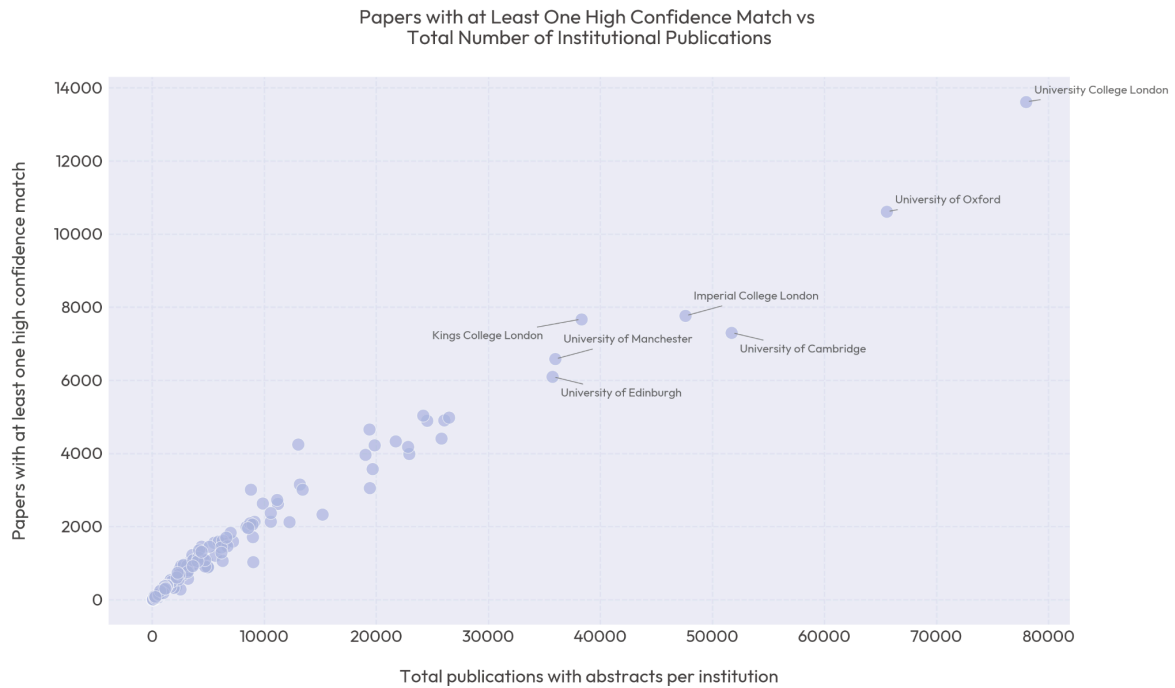

**Figure 5. Number of papers with at least one high confidence match against total institutional publication volume, Overton Engage, 2020-2025:** Each point represents a UK higher education institution. The x-axis shows the total number of publications with abstracts per institution retrieved via OpenAlex; the y-axis shows the absolute number of papers from that institution with at least one high confidence match to a policy engagement opportunity in Overton Engage. Only institutions with more than 30,000 publications with abstracts are labelled.

Figure 5 shows the absolute number of papers with at least one high confidence match against total publication volume. A strong positive linear relationship is visible across the full range of institution sizes suggesting that, broadly speaking, size matters: if a university publishes lots of papers, then it typically (without controlling for subject area) also has a lot of papers relating to policy opportunities.

In this dataset, papers are matched to opportunities on a many-to-many basis, multiple papers may match to the same opportunity. Across all 5,059 UK opportunities analysed, 97.3% have at least one high-confidence match with a paper from a UK institution.

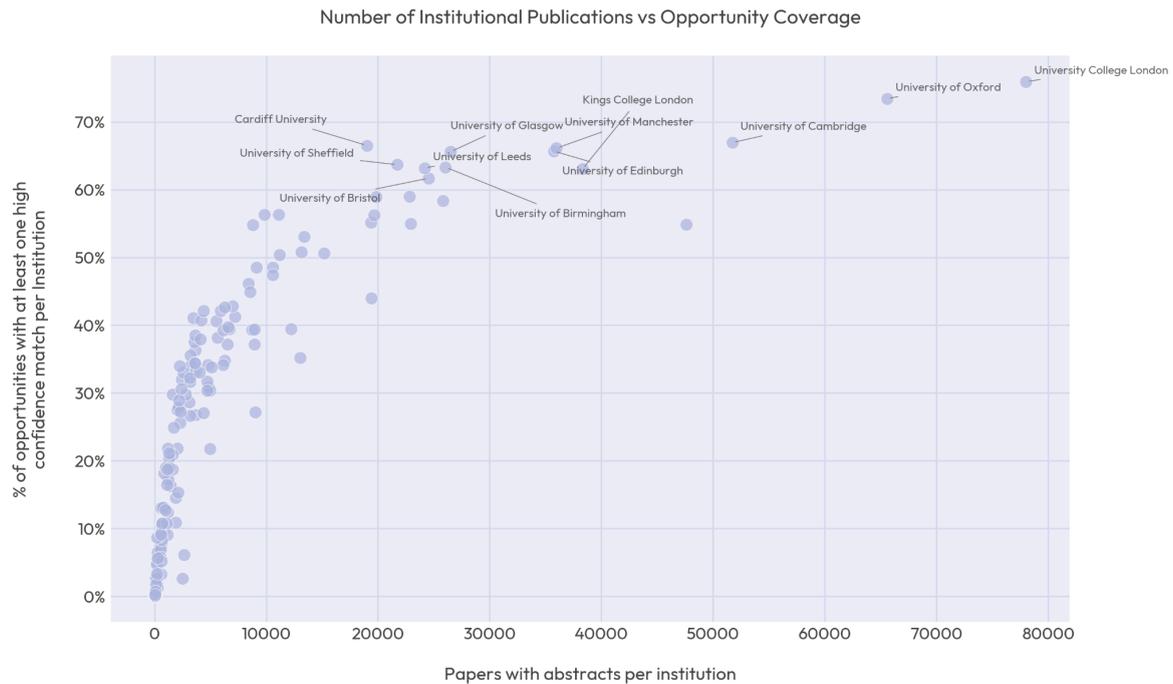

**Figure 6. Institutional publication volume and percentage of policy engagement opportunities with at least one high confidence match, Overton Engage, 2020-2025:** Each point represents a UK higher education institution. The x-axis shows the number of publications with abstracts per institution retrieved via OpenAlex; the y-axis shows the percentage of UK policy opportunities in Overton Engage for which that institution has at least one high confidence match. Only institutions with more than 60% coverage are labelled.

Figure 6 looks at how well each university "covers" the set of opportunities: the percentage of UK policy opportunities for which an institution has at least one high confidence match, plotted against the publication volume. Rather than a simple linear relationship, it follows a saturation style curve, with coverage increasing steeply at lower publication volumes before plateauing among the largest institutions. Larger, more research-intensive institutions tend to have high confidence matches across a greater proportion of opportunities. University College London achieves the highest coverage at around 75%, followed by the Universities of Oxford and Cambridge.

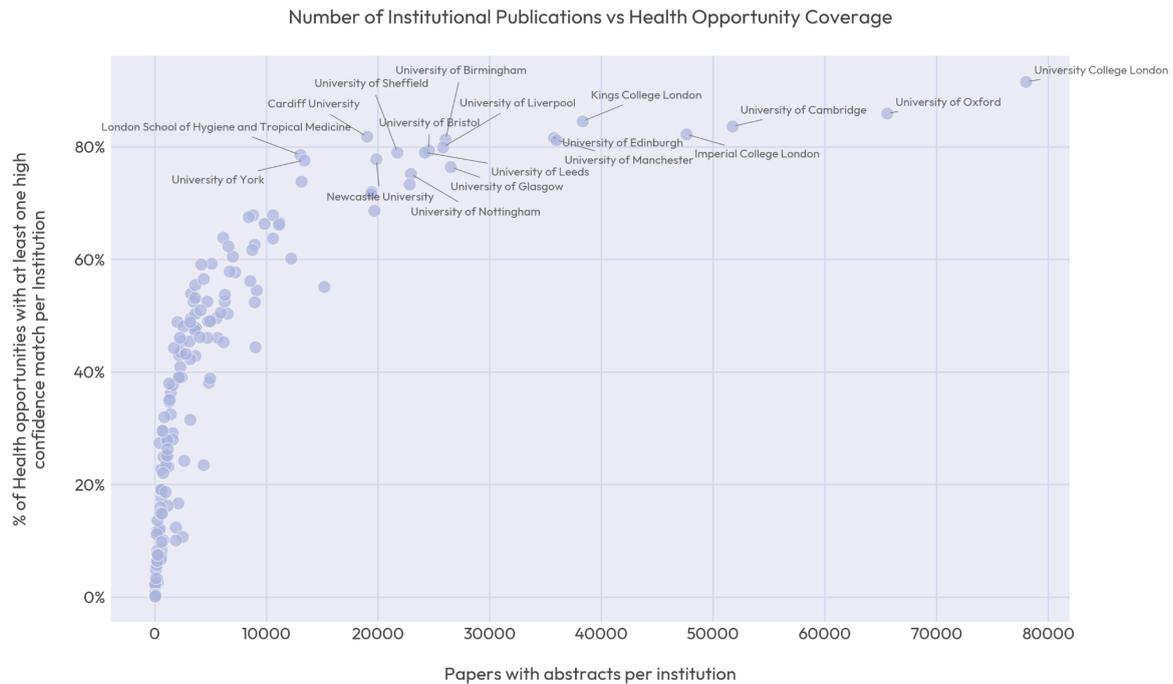

**Figure 7. Institutional publication volume and percentage of health policy engagement opportunities with at least one high confidence match, Overton Engage, 2020-2025:** Each point represents a UK higher education institution. The x-axis shows the number of publications with abstracts per institution retrieved via OpenAlex; the y-axis shows the percentage of UK health policy opportunities in Overton Engage for which that institution has at least one high confidence match. Only institutions with more than 75% coverage are labelled.

Figure 7 restricts the analysis to health opportunities only. Coverage is higher across the board compared to the all-opportunities plot, possibly reflecting both the concentration of UK policy engagement activity in health-related domains and the breadth of health research across the UK university sector. The plot helps to reveal where institutional focus is: a number of smaller, specialist institutions (for example, the London School of Hygiene & Tropical Medicine) achieve higher health opportunity coverage comparable to much larger ones.

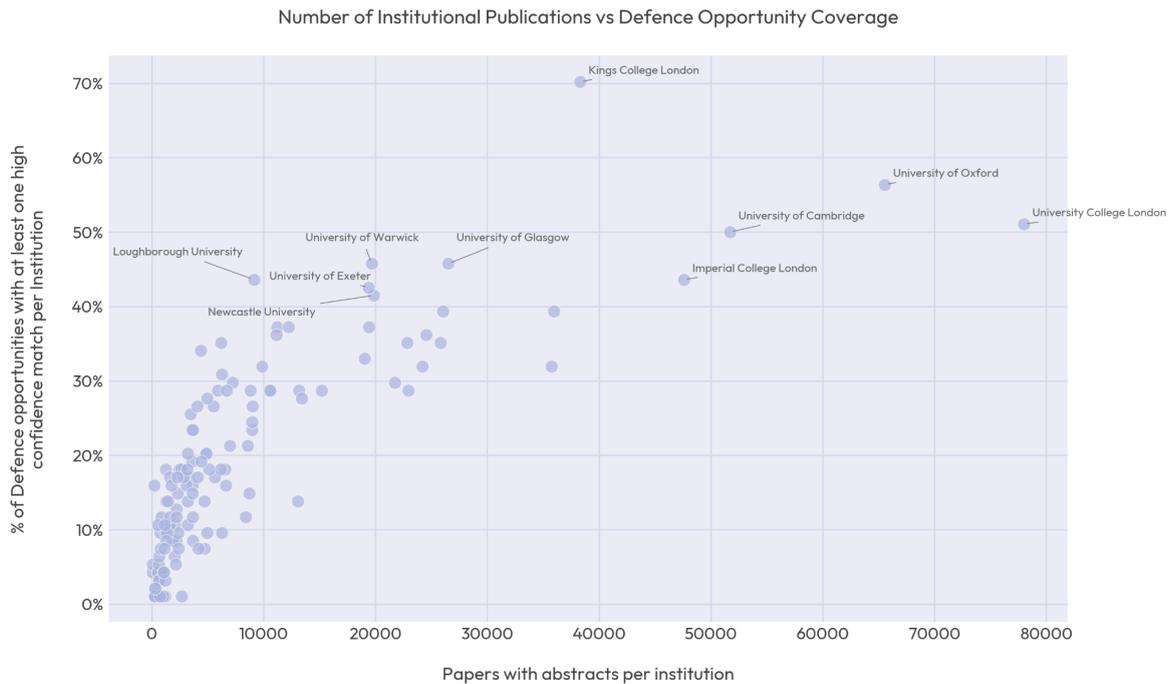

**Figure 8. Institutional publication volume and percentage of defence policy engagement opportunities with at least one high confidence match, Overton Engage, 2020-2025:** Each point represents a UK higher education institution. The x-axis shows the number of publications with abstracts per institution retrieved via OpenAlex; the y-axis shows the percentage of UK defence policy opportunities in Overton Engage for which that institution has at least one high confidence match. Only institutions with more than 40% coverage are labelled.

Figure 8 restricts the analysis to defence opportunities only. Overall coverage is markedly lower than in the health plot, reflecting the more specialised nature of defence policy engagement and the narrower range of institutions with relevant research output. King's College London is a clear outlier, reaching around 70% coverage, notably higher than UCL, Oxford, and Cambridge, which all sit closer to 50%. This may reflect KCL's well-established defence and security research profile (https://www.rand.org/randeurope/about/news/2022/rand-europe-supports-uk-defence-academy.html). Loughborough University's appearance among the labelled institutions is notable, as it is considerably smaller than the other labelled institutions, pointing to the role of specialised engineering and technology research in driving defence policy engagement.

Unlike the coverage plots (Figures 6-8), where the relationship between publication volume and match rate plateaus among larger institutions, the relationship in Figure 5 is broadly linear, indicating that larger institutions continue to accumulate policy-relevant papers at a roughly constant rate relative to their overall output. The spread of points around the trend widens at higher publication volumes, suggesting some variation in how efficiently larger institutions convert research output into policy-relevant papers.

The number of matches to opportunities increases with the volume of papers an institution publishes. However, the number of opportunities with at least one match does not scale linearly with publication output. When examining specific topics, smaller universities can achieve levels of coverage comparable to larger institutions.

**Conclusion**

This paper has introduced Overton Engage, a structured database of publicly documented academic policy engagement opportunities, together with a semantic matching approach that links these opportunities to academic research outputs. By centralising and standardising information that is otherwise fragmented across institutions, platforms, and policy organisations, the system lowers practical barriers to engagement and enables systematic descriptive and comparative analysis of how governments publicly invite academic input.

Engagement opportunities provide a complementary perspective on government policy priorities, reflecting not only what governments publish, but where they actively seek external expertise. All results should therefore be interpreted as describing documented engagement practices rather than the full universe of policymaker-researcher interaction.

In this paper, we have demonstrated how a matching system may be used to assess how effectively individual universities align with domestic policy demand in the UK context. With broader and more balanced international coverage, this approach could be extended to enable comparative analysis of research capacity across countries, identify gaps between policy needs and available expertise, and investigate how different policy systems prioritise and structure academic engagement across disciplines and national research systems.

More broadly, this type of data collection may make it easier to study inequalities in access to engagement opportunities, the relationship between policy demand and research supply, and the role of institutional priorities in shaping knowledge mobilisation.

Several immediate opportunities for future work follow directly from these findings. Expanding geographic and institutional coverage, together with greater consistency in how engagement opportunities are publicly documented, would reduce visibility bias and move the database closer to reflecting the true landscape of academic-policy engagement. Achieving this will require not only technical development, but also greater coordination between governments, policy organisations, and research institutions in how engagement opportunities are published, described, and archived.

Overton Engage is available at [app.overton.io/ui/opportunities](app.overton.io/ui/opportunities). The database can be freely browsed after creating a login and the matching functionality is available with an institutional subscription. We welcome collaboration from researchers, policymakers, and institutions interested in improving the visibility, inclusiveness, and effectiveness of academic policy engagement.

**Supplemental Information**

Opportunity type definitions:

- Calls for evidence and consultations: Organisations may ask for your input on specific topics, often through answering questions or attending meetings.
- Government research priorities: Sometimes known as ARIs in the UK or Learning Agendas in the US, these are broader research questions that steer a government department's work over several years. You might engage by writing a short briefing or attending a meeting.
- Fellowships: These offer a chance to work closely with policymakers on either long-term issues or specific policy areas, typically with hands-on involvement in the policy process.
- Internships: Aimed at early-career researchers, these opportunities let you work within a government department for a few months, gaining first-hand experience in policy making.
- Funding: We also highlight funding opportunities designed to support research that informs and influences policy decisions.
- Expert advisory committees: These opportunities involve joining panels of subject matter experts to provide advice on specific policy challenges.

Table S1: Data Characterisation:

| Institution | Number of Publications | Number of Papers with Abstracts | Number of Papers Matching at Least One Opportunity |
| --- | --- | --- | --- |
| University College London | 118329 | 78008 | 13618 |
| University of Oxford | 98007 | 65570 | 10612 |
| University of Cambridge | 77334 | 51721 | 7304 |
| Imperial College London | 72726 | 47596 | 7762 |
| King's College London | 57823 | 38312 | 7674 |
| University of Manchester | 55762 | 35973 | 6581 |
| University of Edinburgh | 52494 | 35745 | 6100 |
| University of Birmingham | 40320 | 26049 | 4905 |
| University of Liverpool | 39543 | 25827 | 4411 |
| University of Glasgow | 38721 | 26508 | 4987 |
| University of Bristol | 36960 | 24547 | 4901 |
| University of Leeds | 36501 | 24189 | 5033 |
| University of Nottingham | 35349 | 22930 | 3982 |
| University of Southampton | 33169 | 22844 | 4185 |

| University | | | |
|---|---|---|---|
| University of Sheffield | 32430 | 21737 | 4338 |
| Queen Mary University of London | 29812 | 19413 | 3059 |
| Newcastle University | 29356 | 19845 | 4223 |
| University of Warwick | 28732 | 19654 | 3570 |
| University of Exeter | 28622 | 19383 | 4657 |
| Cardiff University | 27485 | 19026 | 3964 |
| Durham University | 20581 | 15185 | 2327 |
| Queen's University Belfast | 20224 | 13160 | 3149 |
| University of York | 19479 | 13399 | 3016 |
| London School of Hygiene and Tropical Medicine | 19092 | 13042 | 4250 |
| University of Leicester | 18770 | 12234 | 2125 |
| Lancaster University | 17227 | 11203 | 2624 |
| University of Surrey | 16810 | 10574 | 2371 |
| University of Strathclyde | 16183 | 11130 | 2732 |
| University of Bath | 15458 | 10587 | 2136 |
| Northumbria University | 14556 | 9856 | 2633 |
| University of Aberdeen | 13984 | 8966 | 1713 |
| Loughborough University | 13970 | 9125 | 2135 |
| London School of Economics | 13638 | 8803 | 3013 |
| Swansea University | 13189 | 8407 | 1989 |
| University of Reading | 13187 | 8955 | 2055 |
| University of Sussex | 12902 | 8556 | 1965 |
| University of St Andrews | 12784 | 9005 | 1030 |

| University | | | |
|---|---|---|---|
| University of East Anglia | 12765 | 8731 | 2094 |
| Manchester Metropolitan University | 10601 | 6985 | 1828 |
| Brunel University of London | 10558 | 6537 | 1529 |
| Liverpool John Moores University | 10517 | 7215 | 1596 |
| University of Ulster | 10293 | 6600 | 1706 |
| University of Essex | 9889 | 6270 | 1612 |
| University of Dundee | 9674 | 6274 | 1063 |
| University of Portsmouth | 9669 | 6173 | 1287 |
| University of Kent | 9642 | 6697 | 1464 |
| Nottingham Trent University | 9334 | 5921 | 1597 |
| University of London | 9124 | 6143 | 1426 |
| Coventry University | 8458 | 5527 | 1559 |
| The Open University | 8243 | 5628 | 1201 |
| University of Plymouth | 8156 | 5115 | 1448 |
| Aston University | 7896 | 4795 | 1039 |
| Royal Holloway, University of London | 7724 | 4947 | 887 |
| Keele University | 7281 | 4950 | 912 |
| Cranfield University | 7273 | 4383 | 1449 |
| University of Hull | 7194 | 4709 | 1081 |
| Heriot-Watt University | 7096 | 4867 | 1075 |
| University of Hertfordshire | 6783 | 4708 | 907 |
| University of Huddersfield | 6679 | 4097 | 1145 |

| University | | | |
|---|---|---|---|
| University of the West of England | 6563 | 4408 | 1312 |
| University of Stirling | 6218 | 4178 | 1356 |
| University of Lincoln | 6089 | 4044 | 1053 |
| Bournemouth University | 6020 | 3694 | 990 |
| Bangor University | 5646 | 3667 | 1062 |
| De Montfort University | 5616 | 3628 | 952 |
| University of Salford | 5527 | 3614 | 919 |
| Sheffield Hallam University | 5422 | 3635 | 1073 |
| Anglia Ruskin University | 5320 | 3258 | 944 |
| University of Greenwich | 5288 | 3647 | 1083 |
| University of Bradford | 5244 | 3100 | 937 |
| Edinburgh Napier University | 5218 | 3583 | 1229 |
| Birmingham City University | 5055 | 3467 | 1092 |
| Birkbeck, University of London | 4870 | 3189 | 580 |
| Oxford Brookes University | 4860 | 3190 | 804 |
| University of Lancashire | 4501 | 3165 | 756 |
| Leeds Beckett University | 4376 | 3202 | 971 |
| University of Derby | 4155 | 2797 | 959 |
| Teesside University | 4043 | 2592 | 933 |
| University of Westminster | 3766 | 2386 | 693 |
| University of Brighton | 3766 | 2332 | 552 |
| Liverpool School of Tropical Medicine | 3763 | 2654 | 668 |

| University | | | |
|---|---|---|---|
| Glasgow Caledonian University | 3588 | 2439 | 861 |
| Middlesex University | 3578 | 2168 | 613 |
| London South Bank University | 3558 | 2305 | 591 |
| University of Wolverhampton | 3344 | 2198 | 605 |
| University of the West of Scotland | 3269 | 2269 | 735 |
| Royal Veterinary College | 3230 | 2494 | 280 |
| Cardiff Metropolitan University | 3218 | 2195 | 581 |
| Kingston University | 3097 | 2047 | 586 |
| SOAS University of London | 3051 | 1896 | 326 |
| Edge Hill University | 3036 | 2042 | 539 |
| Goldsmiths University of London | 3028 | 2125 | 363 |
| Aberystwyth University | 2930 | 1901 | 335 |
| University of South Wales | 2794 | 1738 | 502 |
| University of East London | 2749 | 1609 | 529 |
| University of Chester | 2585 | 1429 | 385 |
| Robert Gordon University | 2502 | 1614 | 553 |
| University of Roehampton | 2414 | 1622 | 364 |
| Canterbury Christ Church University | 2217 | 1453 | 350 |
| University of Winchester | 2184 | 1194 | 325 |
| University of Staffordshire | 2108 | 1337 | 386 |

| Institution | | | |
|---|---|---|---|
| London Metropolitan University | 1959 | 1234 | 404 |
| University of Bedfordshire | 1958 | 1296 | 428 |
| Scotland's Rural College | 1865 | 1169 | 337 |
| University of Sunderland | 1837 | 1290 | 381 |
| York St John University | 1726 | 1146 | 310 |
| University of West London | 1664 | 1105 | 370 |
| University of Gloucestershire | 1619 | 1204 | 377 |
| University of Northampton | 1546 | 1028 | 303 |
| University of Worcester | 1540 | 1066 | 288 |
| London Business School | 1454 | 752 | 222 |
| University of the Arts London | 1432 | 993 | 188 |
| Liverpool Hope University | 1352 | 789 | 210 |
| University of Greater Manchester | 1282 | 854 | 289 |
| University of Suffolk | 1196 | 775 | 194 |
| Queen Margaret University | 1055 | 761 | 236 |
| University of Buckingham | 1050 | 609 | 119 |
| University of the Highlands and Islands | 1039 | 708 | 235 |
| Bath Spa University | 935 | 635 | 172 |
| University of Chichester | 900 | 645 | 154 |
| Harper Adams University | 885 | 593 | 141 |

| University | | | |
|---|---|---|---|
| University of Cumbria | 872 | 640 | 195 |
| Abertay University | 871 | 613 | 154 |
| University of Wales Trinity Saint David | 851 | 591 | 158 |
| University College Birmingham | 838 | 598 | 135 |
| City St. George's, University of London | 787 | 429 | 125 |
| University of Wales | 772 | 540 | 128 |
| Southampton Solent University | 753 | 530 | 142 |
| Royal College of Art | 752 | 533 | 100 |
| St Mary's University, Twickenham | 714 | 518 | 71 |
| Leeds Trinity University | 649 | 426 | 124 |
| Buckinghamshire New University | 521 | 357 | 94 |
| Bishop Grosseteste University | 449 | 255 | 31 |
| Regents University London | 423 | 267 | 82 |
| Wrexham University | 421 | 285 | 77 |
| University for the Creative Arts | 389 | 235 | 44 |
| Hartpury University | 363 | 278 | 41 |
| Glasgow School of Art | 349 | 215 | 53 |
| Falmouth University | 342 | 238 | 25 |
| Arden University | 335 | 214 | 84 |
| Royal Agricultural University | 300 | 204 | 77 |
| University of Law | 296 | 216 | 94 |

| Newman University | 252 | 181 | 46 |
| Plymouth Marjon University | 248 | 181 | 42 |
| BPP University | 233 | 172 | 39 |
| Courtauld Institute of Art | 182 | 127 | 3 |
| Leeds Arts University | 167 | 123 | 24 |
| Royal College of Music | 136 | 113 | 23 |
| Health Sciences University | 136 | 90 | 16 |
| Royal Conservatoire of Scotland | 135 | 94 | 8 |
| Ravensbourne University London | 123 | 75 | 22 |
| Northeastern University London | 108 | 70 | 9 |
| Arts University Bournemouth | 91 | 59 | 5 |
| Richmond American University London | 71 | 55 | 7 |
| Royal Northern College of Music | 71 | 60 | 9 |
| Guildhall School of Music and Drama | 58 | 42 | 4 |
| Leeds College of Music | 53 | 44 | 2 |
| Norwich University of the Arts | 50 | 31 | 5 |
| Royal Academy of Music | 46 | 40 | 2 |
| Arts University Plymouth | 19 | 14 | 1 |
| BIMM University | 16 | 14 | 1 |

Table S2: GPT-4o-mini evaluation scores by match confidence tier:

|  | Green | Yellow | Orange | Red |
| --- | --- | --- | --- | --- |
| Top score (2,2,1) | 44.1% | 26.4% | 21.0% | 12.9% |
| All dimensions positive (Rel≥1, Exp≥1, Scope=1) | 87.5% | 75.3% | 68.5% | 55.0% |
| At least partially relevant (Rel≥1) | 97.7% | 92.8% | 89.7% | 81.6% |
| Not relevant (Rel=0) | 2.3% | 7.2% | 10.3% | 18.4% |

==== Info - methods for the university analysis in this paper

**Publication data**

Publication records for UK higher education institutions were retrieved from OpenAlex (Priem et al., 2022), an open bibliographic database indexing over 250 million scholarly works. For each institution, we queried all works affiliated with that institution published within a five-year window (2020-2025), using OpenAlex institutional identifiers. Works were restricted to the following document types: journal articles, book chapters, books, reports, editorials, and letters; paratext and retracted publications were excluded. Abstracts shorter than 20 characters were treated as missing. Works without usable abstracts were excluded from the matching procedure but retained for publication volume statistics. A compiled list of UK higher education institutions is provided in the Supplementary Material.

**Opportunity preprocessing**

Policy engagement opportunities in Overton Engage vary considerably in format, length, and register, reflecting differences in how government departments and public bodies document and advertise engagement activities. To improve semantic comparability with academic publication abstracts, each opportunity was preprocessed using GPT-4o (OpenAI, 2024). The model was prompted to rewrite each opportunity description into a standardised, research-oriented format: translating administrative and policy language into academic terminology, extracting implicit research questions, and producing a structured output including a rewritten title, background, research questions, keywords, and COFOG classification. This step serves to homogenise the diverse source formats and to bridge the register gap between policy language and scholarly discourse, thereby improving the quality of downstream embedding-based matching.

**Embedding model and fine-tuning**

The pre-processed opportunity descriptions produced in the previous step and publication title-abstract pairs were encoded as dense vector representations using a sentence-transformer model fine-tuned from GTE-Large (Li et al., 2023), a general-purpose text embedding model producing 1024-dimensional vectors. To distinguish between the two text types during encoding, opportunity texts were prefixed with [OPPORTUNITY] and publication texts with [SCHOLAR]. The base model was fine-tuned to improve alignment between policy opportunity descriptions and scholarly works.

Training data for fine-tuning were generated using GPT-4o-mini (OpenAI, 2024). For each opportunity-publication pair in a reference dataset, the model was prompted to evaluate alignment between the original opportunity description and the publication along three dimensions: (1) relevance of the research to the policy question, rated on a three-point scale (not relevant, partially relevant, fully relevant); (2) whether the authors possess the expertise to address the question; and (3) whether the scale and scope of the research align with those of the opportunity. These multi-dimensional scores were combined into a composite relevance score on a continuous scale from 0 to 1 and used as training labels. The embedding model was then fine-tuned on pre-processed (rewritten) opportunity descriptions paired with publication title-abstract pairs, using CoSENT loss (Huang et al., 2024), a ranking loss that optimises pairwise cosine similarities to preserve the ordering of graded similarity labels. Multiple fine-tuning iterations were evaluated against held-out opportunity-publication pairs scored by the same LLM evaluation procedure; the best-performing variant was selected for use in the matching system. The model was run in half-precision (FP16) on GPU hardware.

**Semantic similarity search**

Similarity between opportunity and publication embeddings was computed using L2 (Euclidean) distance. For each opportunity, the 100 most similar publication abstracts were retrieved using FAISS (Johnson et al., 2019; Douze et al., 2024), a library for efficient nearest-neighbour search, with GPU acceleration on an NVIDIA GPU (RTX A4000).

**Match confidence classification**

Each opportunity-publication pair in the top-100 results was assigned a confidence category based on L2 distance thresholds. To calibrate these thresholds, a random sample of 25,000 opportunity-publication pairs from a reference institution was independently evaluated by GPT-4o-mini and manually spot-checked on three dimensions: relevance (0-2), author expertise (0-2), and scale/scope alignment (0-1). Distance thresholds were selected by examining the distribution of evaluation scores across distance bands and choosing cut-points that produced a monotonic quality gradient across tiers. The resulting tiers are: Green (high confidence, L2 ≤ 0.288), Yellow (good, L2 ≤ 0.309), Orange (fair, L2 ≤ 0.334), and Red (weak, L2 ≤ 0.39). Pairs exceeding the Red threshold were excluded. Validation confirmed that match quality degrades consistently across tiers: the proportion of pairs rated positively on all three evaluation dimensions was 87.5% for Green, 75.3% for Yellow, 68.5% for Orange, and 55.0% for Red, while the proportion rated as not relevant rose from 2.3% (Green) to 18.4% (Red) (Table S2). Red-tier matches therefore represent lower-confidence rather than irrelevant matches, roughly half retain some relevance, but the noise level at this tier is considered too high for the purposes of this analysis. Accordingly, the analyses presented here use only Green-tier matches as the primary indicator of high-confidence alignment. The remaining matches were then linked to their authors via OpenAlex authorship records, and authors were ranked by both the frequency of their works in the matched set and the L2 distances of those matches.

======